\documentclass[proof]{WileyASNA-v1}
\usepackage[T1]{fontenc}
\usepackage{ae,aecompl}
\usepackage{graphicx}	
\usepackage{amsmath}	
\usepackage{amssymb}	

\articletype{Original paper}

\received{31 December 2018}
\revised{31 December 2018}
\accepted{31 December 2018}

\begin{document}

\title{Stability of a force-free Hall equilibrium and release of magnetic energy}

\author[]{L.\,L.~Kitchatinov}

\authormark{L.\,L.~Kitchatinov}


\address[\empty]{\orgdiv{ } \orgname{Institute of Solar-Terrestrial Physics}, \orgaddress{\state{P.O.Box\,291, Irkutsk 664033}, \country{Russia}}}

\corres{\email{kit@iszf.irk.ru}}


\abstract{Conservation of magnetic helicity by the Hall drift does not prevent Hall instability of helical fields. This conclusion follows from stability analysis of a force-free spatially-periodic Hall equilibrium. The growth rates of the instability scale as $\sigma \propto B^{3/4}\eta^{1/4}$ with the field strength $B$ and magnetic diffusivity $\eta$ and can be large compared to the rate of resistive decay of the background field. The instability deviates the magnetic field from the force-free configuration. The unstable eigenmodes include a fine spatial structure which evolves into current sheets at the nonlinear stage of the instability. The instability catalyses the resistive release of magnetic energy. The energy is released in a sequence of spikes, every spike emits  several percent of the total energy. A numerically defined scaling for the energy released in a single spike permits an extrapolation to astrophysically relevant values of the Hall number. The instability can be relevant to magnetic energy release in a neutron star crust and, possibly, in stellar coronae.}


\maketitle
\section{Introduction}\label{introduction}
The significance of the Hall effect in astrophysical plasmas can be quantified by the ratio
\begin{equation}
    R_H = \omega_e/\nu_{ei}
    \label{1}
\end{equation}
of the electron gyrofrequency $\omega_e$ to the frequency $\nu_{ei}$ of their collisions with particles of other species \citep[cf., e.g.,][]{S12}. The ratio increases with the magnetic field. Accordingly, the importance of the Hall effect for neutron stars with their strongest known magnetic fields has been long recognised (\citeauthor{J88}, \citeyear{J88}; \citeauthor{GR92}, \citeyear{GR92}). The effect can be significant also for the solar corona \citep{VCO00} or cool protostellar disks (\citeauthor{W99}, \citeyear{W99}; \citeauthor{RK05}, \citeyear{RK05}) where the frequency of electron scattering is relatively low.

The Hall effect does not change the magnetic energy. Paradoxically, the effect is studied in relation to the magnetic energy release. This is because of the ability of the Hall drift to change the field patterns towards small-scale structures thus catalyzing the Ohmic dissipation. \cite{VCO00} showed that the Hall drift in density-stratified plasmas can produce current sheets and cause rapid dissipation of magnetic fields. \cite{GR92} suggested that the magnetic energy cascade due to a hypothetical Hall instability can explain the decrease in pulsar's magnetic fields on a short time scale compared to the diffusive time. \cite{RG02} showed that Hall equilibria with sufficiently strong spatial inhomogeneity are unstable and the instability tends to produce small-scale magnetic structures \citep{GR02}. The instability is driven by the Hall drift but finite magnetic diffusion is necessary  \citep{GH16}.

The Hall instability is not related to a certain geometry or boundary conditions. The stars' spherical geometry or boundaries can therefore be considered as unnecessary complications. A model of a magnetic field periodically varying in space \citep{K17} shows the Hall instability with large growth rates compared to the rate of Ohmic decay. The model's simplicity permitted the computation of the nonlinear dynamics of the instability. The computations show the release of magnetic energy in a sequence of spikes similar to pulsar's bursts or solar flares. This paper extends the model of \cite{K17} to the case of a force-free helical magnetic field.

The motivation for the extension is as follows. Pulsar's magnetic fields are expected to be helical \citep{BS04}. The expectation comes from the fact that purely toroidal or poloidal fields in radiative cores of pulsar progenitors are unstable to interchange disturbances but a helical combination of poloidal and toroidal fields can be stable \citep{W73}. \cite{BS04} have shown that arbitrary initial fields in their numerical simulations eventually relaxed to a helical configuration. Magnetic fields in low $\beta$ plasma of the solar corona are also believed to be force-free and therefore helical \citep[cf., e.g.,][]{P79}. The minimum magnetic energy for a given helicity corresponds to (linear) force-free fields \citep{W58,BN08}. The energy of such fields cannot, therefore, be released via a process conserving helicity. The Hall effect does not change magnetic helicity. Hall instability of helical fields is nevertheless possible \citep{GH16} because of the finite resistivity it requires.

This paper concerns the Hall instability using the magnetic field induction equation alone. The matter's motion is disregarded. The approach is justified for the solid crusts of neutron stars. The force-free initial state is also an appropriate initial equilibrium for stellar coronas. It will therefore be checked whether the instability deviates the field from the force-free state. The simplicity of the model allows computation of the nonlinear stage of the instability. This can be done for a quite large Hall parameter ($\leq 500$) but it will still be much smaller compared with its astrophysically relevant values. The accessible Hall parameters are however large enough for estimating a power-law scaling for extrapolation to still larger $R_H$.

The rest of the paper is organized as follows. The next Sect.\,\ref{s2} recalls the essence of the Hall effect, formulates the magnetic induction equation and specifies the initial equilibrium for the subsequent stability analysis. Section\,\ref{s3} concerns the linear stability and Sect.\,\ref{s4} - nonlinear stability. Finally, Sect.\,\ref{s5} discusses the results and concludes.
\section{Induction equation with Hall effect}\label{s2}
When the Hall parameter (\ref{1}) is not small, electric conductivity is anisotropic and the electric current $\mathbf{j}$ is no longer parallel to the electric field $\mathbf{E}$ \citep{S12}:
\begin{equation}
    \mathbf{j}' = \frac{\sigma_\parallel}{1 + R_H^2}
    \left[\mathbf{E}'
    + R_H\hat{\mathbf{b}}\times\mathbf{E}'
    + R_H^2(\hat{\mathbf{b}}\cdot\mathbf{E}')\hat{\mathbf{b}}
    \right],
    \label{2}
\end{equation}
where primes mean that the equation is formulated for the reference frame moving with the local mass velocity, $\hat{\mathbf b}$ is the unit vector along the magnetic field $\mathbf B$ and $\sigma_\parallel = e^2n_e/(m_e\nu_{ei})$ is the conductivity along the magnetic field. The reversed Eq.\,(\ref{2})
\begin{equation}
    \mathbf{E}' = \sigma_\parallel^{-1}\left( \mathbf{j}' - R_H\hat{\mathbf b}\times\mathbf{j}'\right)
    \label{3}
\end{equation}
shows the canonical Hall effect of the EMF generated by the electric current across the magnetic field.

The induction equation of magnetohydrodynamics with allowance for the Hall effect can be found with the standard procedure \citep[cf. Chap.\,4 in][]{P79} of the transformation to the laboratory frame neglecting the terms of higher than the first order in the ratio $v/c$ of the mass velocity to the velocity of light:
$\mathbf{E}' = \mathbf{E} + \mathbf{v}\times\mathbf{B}/c,\ \mathbf{j}' = \mathbf{j}$. In this approximation, the displacement current in the Maxwell equations is neglected to lead to the Ampere law
\begin{equation}
    \mathbf{j} = c\left(\mathbf{\nabla}\times\mathbf{B}\right)/(4\pi) .
    \label{4}
\end{equation}
Equation (\ref{3}) can then be rewritten as
\begin{equation}
    \mathbf{E} = -\mathbf{v}\times\mathbf{B}/c
    + c(\mathbf{\nabla}\times\mathbf{B})/(4\pi\sigma_\parallel)
    + \mathbf{j}\times\mathbf{B}/(c e n_e).
    \label{5}
\end{equation}
Substitution of this equation into the Maxwell equation $\partial\mathbf{B}/\partial t = - c \mathbf{\nabla}\times\mathbf{E}$ then gives
\begin{equation}
    \frac{\partial \mathbf{B}}{\partial t} = \mathbf{\nabla}\times \left[
    \left(\mathbf{v} + \mathbf{v}_c\right)\times\mathbf{B}
    - \frac{c^2}{4\pi\sigma_\parallel}\mathbf{\nabla}\times\mathbf{B}\right]\ ,
    \label{6}
\end{equation}
where
\begin{equation}
    \mathbf{v}_c = -\mathbf{j}/(e n_e)
    \label{7}
\end{equation}
is the electron current velocity.

Equation (\ref{2}) shows that conductivity across the magnetic field is small in the case of a large Hall number. It may therefore be expected that the current across the magnetic field is also small in this case. Equation (\ref{4}) however shows that the current is controlled solely by the magnetic field configuration. The current across the magnetic field can be not small if $\mathbf{\nabla}\times\mathbf{B}$ has an appreciable cross-component. Induction equation (\ref{6}) resolves the seeming contradiction by showing that the electric current does not cross the magnetic field lines but transports  the field with the current velocity (\ref{7}).

The Hall equilibria are conventionally defined as the magnetic fields nullifying the contribution of the Hall drift in the induction equation (\ref{6}). Obviously, any force-free field $\mathbf{j}\times\mathbf{B} = 0$ makes up such an equilibrium. This paper concerns the stability of a particular force-free equilibrium,
\begin{equation}
    \mathbf{B}_\mathrm{eq} = B_0\left[\hat{\mathbf{y}}\sin (\kappa x) + \hat{\mathbf{z}}\cos (\kappa x)\right] ,
    \label{8}
\end{equation}
to small disturbances. In this equation, $\hat{\mathbf{y}}$ and $\hat{\mathbf{z}}$ are unit vectors along the $y$ and $z$ axes of the Cartesian coordinate system used in this paper. The matter's motion is neglected, $\mathbf{v} = 0$.

From now on, all variables are normalised to dimensionless units with distances measured in units of $\kappa^{-1}$ and time - in units of the diffusive time-scale  $4\pi\sigma_\parallel/(c\kappa)^2$. The same notations are used for dimensionless variables as used before for their physical counterparts. The starting equation for the stability analysis reads
\begin{equation}
    \frac{\partial \mathbf{B}}{\partial t} =
    R_H \mathbf{\nabla}\times\left[\mathbf{B}\times(\mathbf{\nabla}\times\mathbf{B})\right]
    + \Delta\mathbf{B},
    \label{9}
\end{equation}
where the background field amplitude $B_0$ of Eq.\,(\ref{8}) is used to normalise the magnetic field and to define the Hall parameter (\ref{1}).
\section{Linear stability}\label{s3}
\subsection{Equations}
Linear stability analysis starts from Eq.\,(\ref{9}) linearised in small disturbance $\mathbf{b}$ of the Hall equilibrium (\ref{8}),
\begin{eqnarray}
    \frac{\partial \mathbf{b}}{\partial t} &=&
    R_H \mathbf{\nabla}\times\left[
    \mathbf{B}_\mathrm{eq}\times(\mathbf{\nabla}\times\mathbf{b})
    + \mathbf{b}\times(\mathbf{\nabla}\times\mathbf{B}_\mathrm{eq})
    \right]
    \nonumber \\
    &+& \Delta\mathbf{b}.
    \label{10}
\end{eqnarray}
It is convenient to separate the disturbance in its poloidal and toroidal parts,
\begin{equation}
    \mathbf{b} = \mathbf{\nabla}\times\left[
    \hat{\mathbf{x}}T + \mathbf{\nabla}\times (\hat{\mathbf{x}} P)\right] ,
    \label{11}
\end{equation}
so that the divergence-free of the field $\mathbf{b}$ is automatically guaranteed.
The equation for the potential $P$ of the poloidal field can be found as an $x$-component of Eq.\,(\ref{10}) after substituting the representation (\ref{11}) into this equation. The $x$-component of the curled Eq.\,(\ref{10}) gives the equation for the potential $T$ of the toroidal field.

Slow Ohmic decay of the background field (\ref{8}) is neglected in this Section and the field $\mathbf{B}_\mathrm{eq}$ is assumed to be steady (the assumption is waved in nonlinear computations of Sect.\,\ref{s4}). With this assumption, coefficients in the equations for magnetic disturbances do not depend on time. They do not depend on $y$ and $z$ coordinates either. The dependence of the form $\exp (ik_2y + ik_3z + \sigma t )$ on these variables can, therefore, be assumed for the disturbances. This reduces the linear stability analysis to the eigenvalue problem:
\begin{eqnarray}
    \sigma P &=& i R_H \left( k_2\sin x + k_3\cos x\right)
    \left( P - T \right) + \frac{\partial^2 P}{\partial x^2} - k^2 P ,
    \nonumber \\
    \sigma T &=& i R_H \left( k_2\sin x + k_3\cos x\right)
    \left( \frac{\partial^2 P}{\partial x^2} - k^2 P + P\right)
    \nonumber \\
    &+& \frac{\partial^2 T}{\partial x^2} - k^2 T ,
    \label{12}
\end{eqnarray}
where $k^2 = k_2^2 + k_3^2$. Coefficients in the Eqs.\,(\ref{12}) vary periodically with $x$. Solution of the equations is therefore periodic as well,
\begin{eqnarray}
    T &=& T_0 + \sum\limits_{n=1}^N \left( T_n^{(+)}\mathrm{e}^{inx}
    + T_n^{(-)}\mathrm{e}^{-inx}\right) ,
    \nonumber \\
    P &=& P_0 + \sum\limits_{n=1}^N \left( P_n^{(+)}\mathrm{e}^{inx}
    + P_n^{(-)}\mathrm{e}^{-inx}\right) .
    \label{13}
\end{eqnarray}
Substitution of (\ref{13}) into (\ref{12}) leads to the eigenvalue problem for a system of $4N + 2$ algebraic equations. The problem was solved numerically. Computations with $N =100$ provide sufficient resolution for $R_H \leq 10^4$.

\begin{figure}[thb]
    \includegraphics[width=\hsize]{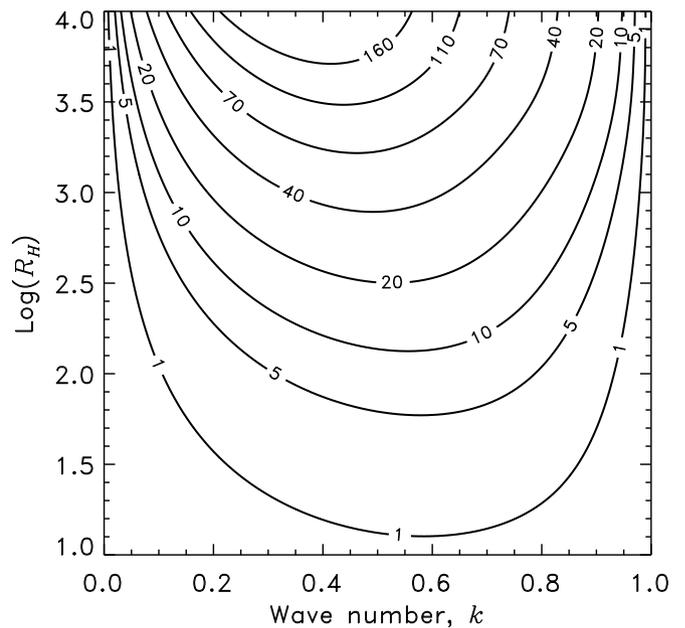}
    \caption{Growth rate isolines on the parametric plane of the Hall
            instability. Numbers in the isoline gaps show the growth rates normalised to the rate $(c\kappa)^2/(4\pi\sigma_\parallel)$ of the Ohmic decay.
            }
    \label{f1}
\end{figure}

The numerical code was written for the Hall equilibrium $\mathbf{B}_\mathrm{eq} = 2B_0\left[q\hat{\mathbf{y}}\sin (\kappa x) + (1-q)\hat{\mathbf{z}}\cos (\kappa x)\right]$, which returns to Eq.\,(\ref{8}) for $q = 0.5$. For $q=1$, it turns into the case considered in \citep{K17} and for $q = 0$ it is equivalent to this case. With these cases, the numerical code was tested and its satisfactory performance confirmed.
\subsection{Results}
The disturbances, which differ in direction of the 2D wave vector $\mathbf{k} = \hat{\mathbf{y}}k_2 + \hat{\mathbf{z}}k_3$ only, are physically equivalent: the disturbances can be converted into each other by a transformation of rotation about the $x$-axis and displacement along this axis. The disturbances growth rates therefore depend on the modulus $k$ of the wave vector $\mathbf{k}$ but not on the direction of this vector.

\begin{figure}[thb]
    \includegraphics[width=\hsize]{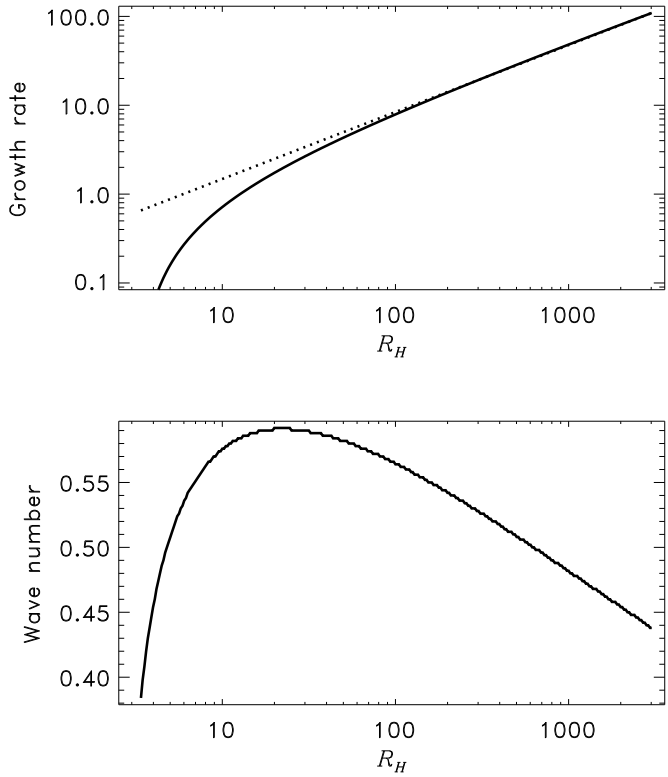}
    \caption{Growth rate of the most rapidly growing instability mode (top panel)
            and the corresponding wave number (bottom panel). The dashed line in the upper panel shows the power-law approximation $\sigma \propto R_H^\gamma$ ($\gamma = 0.753$).
            }
    \label{f2}
\end{figure}

Hall instability of the force-free equilibrium (\ref{8}) is non-oscillatory. All unstable eigenmodes have (positive) real eigenvalues.  Figure~\ref{f1} shows the stability map. The large growth rates justify the neglect of the slow diffusive decay of background field.

Figure~\ref{f2} shows the dependence of the growth rate and the wave number of the most rapidly growing disturbance on the Hall parameter. For large Hall number, the growth rate of Fig.\,\ref{f2} approaches the power law $\sigma \propto R_H^\gamma$. The power index is $\gamma = 3/4$ within the numerical error. This value shows that finite conductivity is necessary for the instability (otherwise the power index would equal one).

\begin{figure}[thb]
    \includegraphics[width=\hsize]{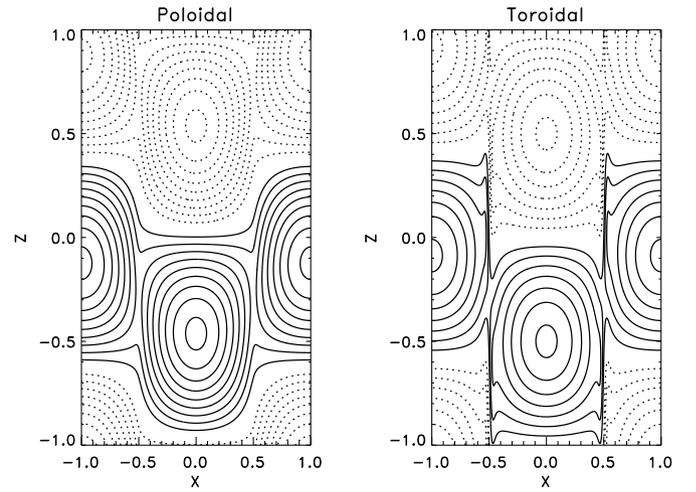}
    \caption{Field lines of the poloidal field (left panel) and isolines
            of the toroidal $y$-component of the field (right) for an unstable eigenmode on the plane of coordinates $X = x/\pi$ and $Z = kz/\pi$. Full (dotted) lines show the clockwise (anti-clockwise) circulation for the poloidal field vector and positive (negative) levels of the toroidal field. The plot shows the most rapidly growing ($\sigma = 21.6$ for $k = 0.552$) disturbance for $R_H = 350$.
            }
    \label{f3}
\end{figure}

The wave numbers of Fig.\,\ref{f2} show that the spatial scales of the instability in the dimensions normal to the $x$-axis are large. However, the eigenmode of Fig.\,\ref{f3} includes small scales in its variation along the $x$-axis. For certainty, the $\mathbf{k}$-vector is aligned with the $z$-axis in this plot and hereafter. The linear eigenmodes are defined up to an uncertain constant factor. The modes were normalised so that the largest coefficient in the series (\ref{13}) equals one in absolute value. The modes are still uncertain up to a factor of $\mathrm{e}^{i\phi}$ after this normalization. The phase $\phi$ was adjusted so that the toroidal $y$-component of the field is zero at the origin $x = z = 0$ of the coordinates.

The unstable disturbances of the helical background field (\ref{8}) are helical as well. The eigenmode of Fig.\,\ref{f3} varies on small scales in the vicinity of $x = \pm0.5\pi$. These are the seeds for current sheets formed at the nonlinear stage of the instability.
\section{Nonlinear dynamics}\label{s4}
\subsection{Equations}
The linear instability modes of the preceding Section are two-dimensional: they vary along the $x$-axis and along the direction of the vector $\mathbf{k}$ normal to this axis. For the wave vector aligned with the $z$-axis, the problem at hand is 2D with homogeneity along the $y$-axis. The 2D magnetic field is then convenient to write as
\begin{equation}
    \mathbf{B} = \hat{\mathbf{y}}B(x,z) + \mathbf{\nabla}\times(\hat{\mathbf{y}} A(x,z)) ,
    \label{14}
\end{equation}
where $A$ is the poloidal field potential. Equations (\ref{14}) and (\ref{9}) then yield
\begin{eqnarray}
    \frac{\partial B}{\partial t} &=& R_H
    \left( \frac{\partial A}{\partial x}\frac{\partial(\Delta A)}{\partial z}
    - \frac{\partial A}{\partial z}\frac{\partial(\Delta A)}{\partial x}\right)
    + \Delta B ,
    \nonumber \\
    \frac{\partial A}{\partial t} &=& R_H
    \left( \frac{\partial B}{\partial x}\frac{\partial A}{\partial z}
    - \frac{\partial B}{\partial z}\frac{\partial A}{\partial x}\right)
    + \Delta A ,
    \label{15}
\end{eqnarray}
where $\Delta = \partial^2/\partial x^2 + \partial^2/\partial z^2$ is the 2D Laplacian. It may be noted that the toroidal field alone is not affected by the Hall drift as it is in spherical geometry  \citep{SU97,Mea14}. The reason is that the poloidal current is aligned with the toroidal field isolines in Cartesian geometry but this is not the case with spherical geometry.

The initial value problem for Eqs\,(\ref{15}) was solved numerically in the domain of $-\pi \leq x \leq \pi ,\ -\pi/k \leq z \leq \pi/k$ with periodic boundary conditions. The initial condition is the superposition of the background field of Eq.\,(\ref{8}) and a small admixture of the most rapidly growing mode of the linear stability problem:
\begin{eqnarray}
    A_0(x,z) &=& \sin x + \varepsilon \Re [ikP(x,z)] ,
    \nonumber \\
    B_0(x,z) &=& \sin x + \varepsilon \Re [ikT(x,z)] ,
    \label{16}
\end{eqnarray}
where $P$ and $T$ are the poloidal and toroidal field potentials of Eq.\,(\ref{11}) and $\varepsilon = 0.01$ in all the computations.

The significance of Hall instabilities is mainly related to their associated release of magnetic energy. The energy normalised to its initial value,
\begin{equation}
    E = \frac{k}{4\pi^2}\int\limits_{-\pi}^\pi\int\limits_{-\pi/k}^{\pi/k}
    \left[ B^2 + \left(\frac{\partial A}{\partial x}\right)^2
    + \left(\frac{\partial A}{\partial z}\right)^2\right]
    \rm{d}z\, \rm{d}x ,
    \label{17}
\end{equation}
and similarly normalised rate of its resistive dissipation,
\begin{equation}
    W = \frac{k}{4\pi^2}\int\limits_{-\pi}^\pi\int\limits_{-\pi/k}^{\pi/k}
    \left[ (\Delta A)^2 + \left(\frac{\partial B}{\partial x}\right)^2
    + \left(\frac{\partial B}{\partial z}\right)^2\right]
    \rm{d}z\, \rm{d}x ,
    \label{18}
\end{equation}
were therefore monitored in the course of the computations. The parameter
\begin{equation}
    F =  \frac{k}{4\pi^2}\int\limits_{-\pi}^\pi\int\limits_{-\pi/k}^{\pi/k}
    \frac{\mathbf{j}\cdot\mathbf{B}}{\sqrt{j^2 B^2 + \epsilon}}\
    \rm{d}z\, \rm{d}x
    \label{19}
\end{equation}
was also monitored to find out whether the magnetic field deviates from the initial force-free state in the course of the instability. The parameter $\epsilon = 10^{-6}$ is inserted in Eq.\,(\ref{19}) to exclude division by zero in numerical integration.

\begin{figure}[thb]
    \includegraphics[width=\hsize]{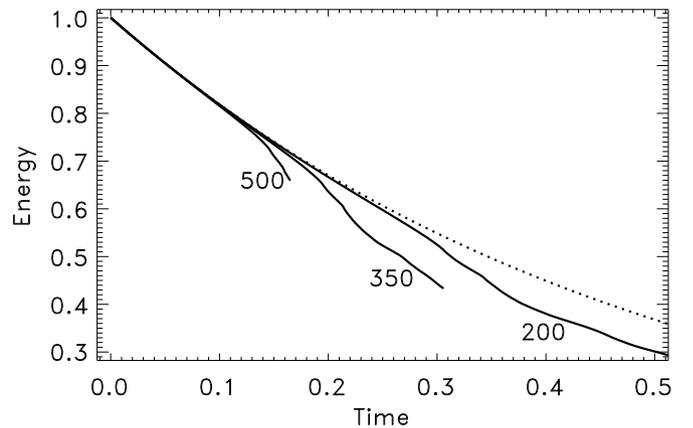}
    \caption{Normalized magnetic energy of Eq.\,(\ref{17}) in nonlinear computations
            of the Hall instability. Different lines are marked by the corresponding values of the Hall parameter $R_H$. The dotted line shows the trend $\exp(-2t)$ for the Ohmic dissipation without the Hall effect ($R_H = 0$).
            }
    \label{f4}
\end{figure}

Equations (\ref{15}) were solved by numerical time-stepping and second-order accurate finite-difference representation of spatial derivatives on a uniform grid of $N = N_x = N_z$ grid points in either spatial dimension. The numerical solution meets a severe resolution problem. The Hall parameter of Eq.\,(\ref{1}) does not include any spatial scale. This means that there is no sufficiently small scale in the problem for which the field dynamics are dominated by diffusion. The standard von Neuman criteria for numerical stability do not apply to the nonlinear problem. Stable operation of the numerical code can be interrupted eventually by an instability. Formerly \citep{K17}, the numerical instability has been possible to avoid or postpone by decreasing the time step. For this reason, the adaptive time-stepping in the fourth-order Runge-Kutta scheme \citep{Pea92} was applied in the computations of this paper. The numerical instabilities due to insufficient spatial resolution were met, nevertheless, but their onset can be delayed by increasing the number $N$ of the grid points. Computations for larger Hall parameter demand higher $N$. The rule $N = \max (200, 2R_H)$ helped to advance the computations for all accessible values of the Hall number beyond the first spike of energy release as discussed in what follows. With this rule, available computational facilities allowed computations up to the Hall number of $R_H = 500$.

\begin{figure}[thb]
    \includegraphics[width=\hsize]{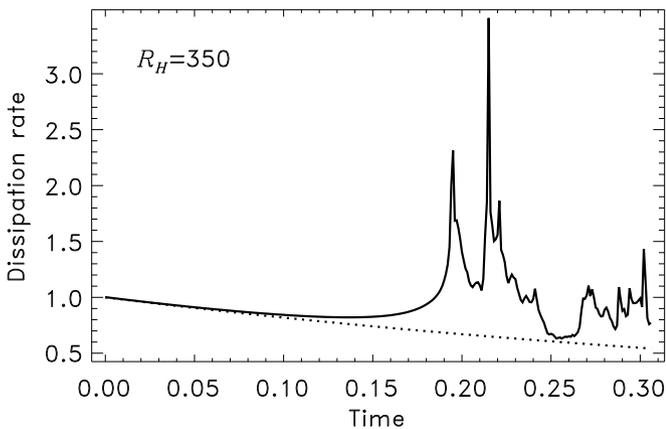}
    \caption{Dissipation rate of Eq.\,(\ref{18}) in the computation for the
             Hall parameter $R_H = 350$. The dotted line shows the Ohmic dissipation rate without the Hall effect ($R_H = 0$).
            }
    \label{f5}
\end{figure}

\subsection{Results}
Hall instabilities can catalyse release of magnetic energy. Figures \ref{f4} and \ref{f5} evidence the enhanced dissipation at the nonlinear stage of the instability. The dotted line in Fig.\,\ref{f4} shows the trend, which the magnetic energy of Eq.\,(\ref{17}) follows without the Hall effect ($R_H = 0$). Computations for finite Hall numbers follow initially the same trend. The dissipation rate (\ref{18}) is increased by the instability upon the time $\tau \approx -\ln (\varepsilon)/\sigma \approx 4.6/\sigma$ when an initially small unstable disturbance grows to a level comparable to the background field. As the growth rate $\sigma$ increases with the Hall number, the energy release caused by the instability starts sooner for larger $R_H$.

\begin{figure}[thb]
    \includegraphics[width=\hsize]{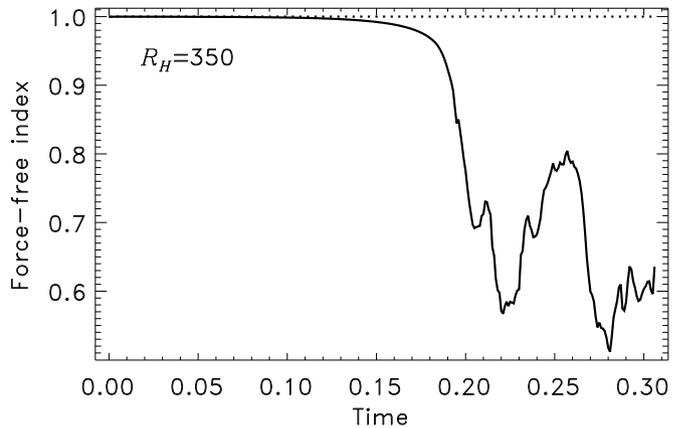}
    \caption{Force-free parameter of Eq.\,(\ref{19}) in the same computation for
             $R_H = 350$ as Fig.\,\ref{f5}.
            }
    \label{f6}
\end{figure}

The dissipation rate profile for the case of $R_H = 350$ is shown in Fig.\,\ref{f5}. The profile consists of a series of spikes where the dissipation rate can exceed several times the rate of Ohmic dissipation of the background field. This type of spiky dissipation is typical for all computations with a sufficiently large Hall number $R_H \geq 200$.

\begin{figure*}[thb]
    \includegraphics[width=\hsize]{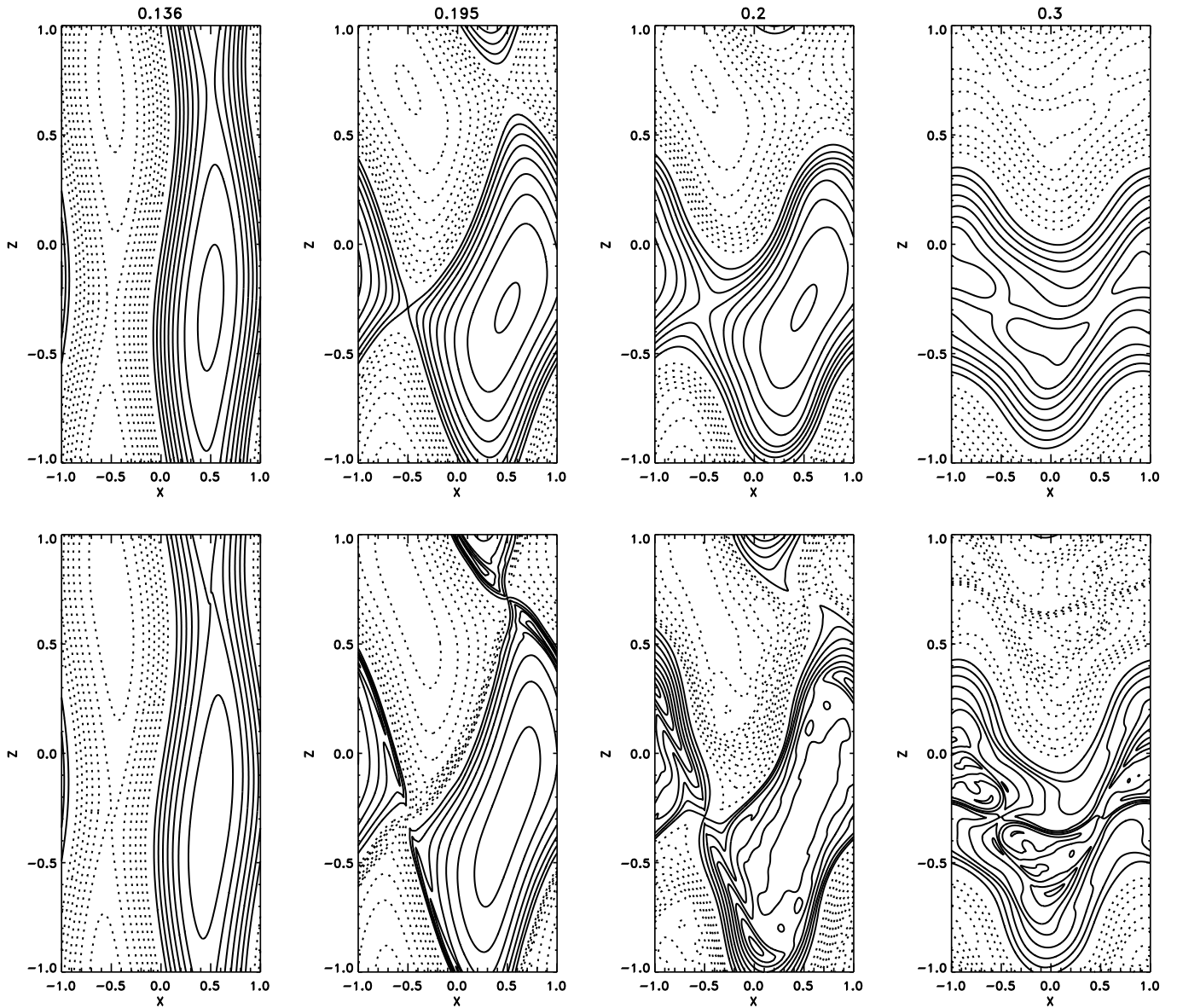}
    \caption{Magnetic field patterns on the plane of the normalised coordinates
            $X = x/\pi$ and $Z = kz/\pi$ for several instants of the computation with
            $R_H = 350$. The upper row shows the poloidal field lines. The full and dotted lines indicate the clockwise and anti-clockwise circulation of the field vector respectively. Normalized time instants corresponding to the individual columns of the Figure are given on the top. The bottom row shows the toroidal file isolines. Full and dotted lines in the bottom row show the positive and negative levels respectively.
            }
    \label{f7}
\end{figure*}

The evolution of the force-free parameter of Fig.\,\ref{f6} shows that the field structure is changed by the Hall instability. The field deviates from the force-free configuration. Figure\,\ref{f7} unfolds the variation of the field pattern in the course of the instability. The first column in this figure corresponds to the instant $t = 0.136$ when the dissipation rate of Fig.\,\ref{f5} passes through its first minimum. The rate increases afterwards. The increase is probably caused by formation of current sheets seen in the poloidal field pattern. Magnetic reconnection in the current sheets increases the dissipation rate. The sheets' length reduces with time and dissipation rate increases contemporarily so that the current sheets transform into the X-points at the instant $t=0.195$ of the first spike of dissipation in Fig.\,\ref{f5}. This stage of the smooth increase in the dissipation rate up to its first spike at the instant of the X-point's formation was common for all computations with not too small $R_H$. After the first spike, however, the fields \lq forget' their initial state. The dynamics after the first spike depends on the particular value of the Hall number. The two last columns in Fig.\,\ref{f7} show the development of Hall turbulence envisaged by \citet{GR92}.

The instability reduces magnetic energy. In the cases when operation of the numerical code was not broken up to $t \simeq 1$, the computations show the decay of the turbulence and relaxation of the field structure to a stable steady state with reduced magnetic energy.

\begin{figure}[thb]
    \includegraphics[width=\hsize]{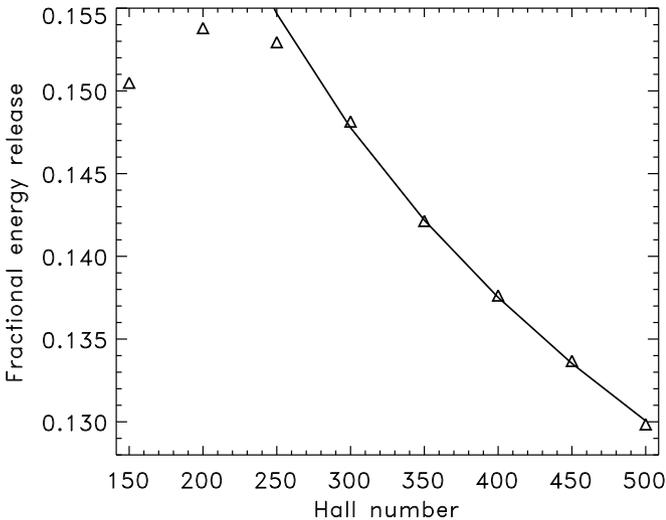}
    \caption{Computed values of the fractional energy release of Eq.\,(\ref{20})
            are shown by triangles. The full line shows the power-law approximation of Eq.\,(\ref{21}).
            }
    \label{f8}
\end{figure}

The largest Hall number $R_H = 500$ accessible for the present computations is still much smaller compared to astrophysical cases. The accessible $R_H$ are however large enough for estimating the scaling relation which can be extrapolated to larger $R_H$. Figure\,\ref{f8} shows the fractional energy release
\begin{equation}
    \frac{\Delta E}{E} = \frac{E(t_\mathrm{min})}{E(t_\mathrm{sp})} - 1
    \label{20}
\end{equation}
between the instants of the first spike ($t_\mathrm{sp}$) and the preceding minimum ($t_\mathrm{min}$) in the dissipation rate as a function of the Hall number
(note that the first spike and the preceding minimum are common for all the computations). For large $R_H$, the dependence is close to the power law
\begin{equation}
    \Delta E/E = 0.615\,R_H^{-1/4} .
    \label{21}
\end{equation}
The fractional energy release, therefore, decreases with the field strength $B$ but its absolute value increases in proportion to $B^{3/4}$.
\section{Discussion}\label{s5}
Conservation of magnetic helicity by the Hall effect does not preclude instability of helical Hall equilibria. This was shown for a particular case of the spatially periodic force-free field. The helicity conservation is violated by finite diffusion which is necessary for the instability \citep{GH16}. The diffusive nature of the Hall instability is evidenced by the scaling
\begin{equation}
    \sigma \propto B^\gamma \eta^{1-\gamma}
    \label{22}
\end{equation}
for the (dimensional) linear growth rates for large Hall numbers ($B$ is the background field amplitude, $\eta$ is the magnetic diffusivity, and $\gamma = 3/4$ in the model of this paper). The analogy with the shear-Hall instability \citep{RS04,RH04} proposed by \citet{K17} is therefore not complete. The analogy helps to explain the origin of the Hall instability and criteria for its onset. The difference however is that shear flow can change the magnetic energy that the shear in the effective velocity (\ref{7}) cannot do. The shear-Hall instability is not diffusive ($\gamma = 1$).

Linear growth rates in this paper do not depend on the direction of the 2D wave vector in the $yz$ plane. This causes an uncertainty in the results. The shape of the most rapidly growing mode is not definite because any superposition of arbitrary number of eigenmodes with different directions of their wave vectors grows with the same rate. The superposition principle does not apply to the nonlinear stage of the instability. Which nonlinear mode wins the race of unstable growth is, therefore, not known. (This is similar to thermal convection in a horizontally unbounded layer, where the linear growth rates do not depend on the direction of the horizontal wave vector, and hexagonal B\'enard cells emerge from a superposition of three linear eigenmodes with different wave vectors in the nonlinear change of stability.)

2D computations of this paper do not necessarily show the most efficient nonlinear mode of magnetic energy release. Nevertheless, the computations unambiguously show that the instability catalyzes the energy release by forming the current sheets. The energy release proceeds in a series of spikes, which are at least qualitatively similar to the bursts of pulsars or stellar flares.

The supposed bursts' energy can be estimated from the scaling of Eq.\,(\ref{21}). \cite{GR92} gave the expression
\begin{equation}
    R_\mathrm{H} \sim 400 \frac{B_{12}}{T_8^2}
    \left(\frac{\rho}{\rho_\mathrm{n}}\right)^2
    \label{23}
\end{equation}
for the Hall number of neutron stars' crust. In this equation, $\rho_\mathrm{n} = 2.8\times10^{14}$ g/cm$^3$ is the \lq nuclear density', $B_{12}$ is the magnetic field in units of $10^{12}$\,G and $T_8$ is the temperature in units of $10^8$\,K. Substitution of this equation into Eq.\,(\ref{21}) gives
\begin{equation}
    \Delta E \sim 0.5 B_{12}^{7/4} L_6^3\sqrt{T_8\rho_\mathrm{n}/\rho}\times 10^{40}\ \mathrm{erg} ,
    \label{24}
\end{equation}
where $L_6$ is the magnetic field scale in units of 10\,km. The estimation is not far from the energies of $\gamma$-ray bursts \citep{H13}.

The estimations for solar corona are also possible \citep{K17} but seem to be premature before stability is analysed with allowance for the fluid motion. Figure\,\ref{f6} shows that the instability deviates the field structure from the force-free state. The fluid will start moving in the course of the instability. It is not clear at the moment how large the fluid inertia should be for the motion being not significant for the Hall instability.

The Hall effect has long been recognised as possibly a significant modification of canonical MHD reconnection models \citep{BP07}. The modification can increase growth rates of the tearing instability \citep{ZMW17}. The above computations however show that the Hall effect alone can produce current sheets and catalyse resistive release of magnetic energy.
\subsection*{Acknowledgments}
This work was supported by the Russian Foundation for Basic Research (project 17-02-00016) and by budgetary funding of the Basic Research program II.16.
\bibliography{Paper}
\end{document}